\documentclass[preprint,notoc]{JHEP3}

\usepackage{epsfig} \usepackage{amssymb}

\newcommand{\eq}[1]{Eq. (\ref{#1})}
\newcommand{\fig}[1]{Fig. \ref{#1}}
\newcommand{\sektion}[1]{Section \ref{#1}}
\newcommand{\ocite}[1]{Ref. \cite{#1}}
\newcommand{\ud}{\mathrm{d}}

\title{Unparticle physics with broken scale invariance}

\author{Vernon Barger$^a$, Yu Gao$^a$, Wai-Yee Keung$^b$,
Danny Marfatia$^c$ and  V. Nefer \c{S}eno\u{g}uz$^c$\\
$^a$Department of Physics, University of Wisconsin, Madison, WI 53706, USA\\
$^b$Department of Physics, University of Illinois, Chicago, IL 60607-7059, USA\\
$^c$Department of Physics and Astronomy, University of Kansas, Lawrence, KS 66045, USA\\
E-mail: \email{barger@pheno.wisc.edu}, \email{yugao@physics.wisc.edu}, \email{keung@uic.edu},
\email{marfatia@ku.edu}, \email{nefer@ku.edu}}

\abstract{
If scale invariance is exact, unparticles are unlikely to be probed in colliders
since there are stringent constraints from astrophysics and
cosmology. However these constraints are inapplicable if scale invariance is broken 
at a scale $\mu\gtrsim1{\rm~GeV}$. The case $1{\rm ~GeV}\lesssim\mu<M_Z$ is particularly
interesting since it allows unparticles to be probed at and below the Z pole.
We show that $\mu$ can naturally be in this range if only vector unparticles
exist, and briefly remark on implications for Higgs phenomenology. We then
obtain constraints on unparticle parameters from $e^+e^-\to\mu^+\mu^-$ cross-section
and forward-backward asymmetry data, and compare with the constraints from mono-photon
production and the $Z$ hadronic width.}

\begin{document}

\section{Introduction} \label{intro}
Unparticle physics was introduced in Ref. \cite{Georgi:2007ek} as a low energy
effective description of a hidden sector with a nontrivial infrared fixed point.
This sector is assumed to interact with the Standard Model (SM) through the exchange
of particles at a high scale $M$. Below $M$, the interactions are of the form
\begin{equation} \label{effop1}
\frac{C_i}{M^{d_{UV}+d_{SM}^i-4}}O^i_{SM}O_{UV}\,,
\end{equation}
where $C_i$ are dimensionless constants, $O^i_{SM}$ is an operator with mass dimension $d_{SM}^i$ built out of SM fields 
and $O_{UV}$ is an operator with mass dimension $d_{UV}$ built out of the hidden sector fields.
Scale invariance in the hidden sector emerges at an energy scale $\Lambda<M$.
In the effective theory below $\Lambda$ the interactions of \eq{effop1} take the form
\begin{equation} \label{effop2}
\frac{C_i\Lambda^{d_{UV}-d}}{M^{d_{UV}+d_{SM}^i-4}}O^i_{SM}O\,,
\end{equation}
where $d$ is the scaling dimension of the unparticle operator $O$.

Unparticle effects might be detectable in missing energy distributions and
interference with SM amplitudes
\cite{Georgi:2007ek,Georgi:2007si,Cheung:2007zz,Cheung:2007ap}.  However, if
scale invariance is exact, unparticles are unlikely to be probed in colliders
since there are strong constraints from astrophysics and cosmology 
\cite{Davoudiasl:2007jr,Freitas:2007ip} (see \sektion{bound}). As discussed in Section \ref{break},
these constraints are inapplicable if scale invariance is broken at a scale
$\mu\gtrsim1{\rm~GeV}$, while constraints from experiments at center-of-mass energy
$\sqrt{s}>\mu$ remain relevant and resonance-like behavior at $\mu$ is expected.  \ocite{Rizzo:2007xr} has
considered collider phenomenology for $\mu>M_Z$.  Here we consider the
constraints on unparticle parameters assuming $1{\rm ~GeV}\lesssim\mu<M_Z$
which allows unparticles to be probed by $s$ channel Z exchange observables. 

For scales of $\Lambda$ and $M$ that are experimentally accessible, the Higgs coupling
to scalar unparticles generally breaks scale invariance at the electroweak scale
\cite{Fox:2007sy,Bander:2007nd}. Having $\mu<M_Z$ in this case requires somewhat small
dimensionless couplings (\sektion{scalaru}). However, if only vector unparticles exist, scale
invariance is broken by higher dimensional operators, and $\mu$ can naturally be below $M_Z$ (\sektion{vectoru}).  We also briefly discuss
how vector unparticles could affect Higgs phenomenology in Section
\ref{higgs}. 

Constraints on vector and axial-vector unparticle couplings obtained using
$e^+e^-\to\mu^+\mu^-$ forward-backward asymmetry (FBA) and total cross-section data
are presented in \sektion{collide}. The resonance-like behaviour at $\mu$
is taken into account in the analysis. Our summary is followed by
four appendices covering details of the unparticle contribution to the
$Z$ hadronic width, the bound from SN 1987A cooling, 
the vacuum polarization correction to the unparticle propagator, and the 
initial state QED corrections.

\section{Bounds on vector unparticle interactions} \label{bound}
Consider vector unparticles coupling to fermions:
\begin{equation} \label{vecop}
{\cal L}_{\psi}=C_V\frac{\Lambda^{d_{UV}-d}}{M^{d_{UV}-1}}\bar{\psi}\gamma_{\mu}\psi O^{\mu}
+C_A\frac{\Lambda^{d_{UV}-d}}{M^{d_{UV}-1}}\bar{\psi}\gamma_{\mu}\gamma_5\psi O^{\mu}
\,,
\end{equation}
which, following the convention of Refs. \cite{Georgi:2007si,Luo:2007me}, can be written as
\begin{equation} \label{veceff}
\frac{c_V}{M_Z^{d-1}}\bar{\psi}\gamma_{\mu}\psi O^{\mu}+
\frac{c_A}{M_Z^{d-1}}\bar{\psi}\gamma_{\mu}\gamma_5\psi O^{\mu}\,,
\end{equation}
with
\begin{equation} \label{convert}
c_{V,A}=C_{V,A}\left(\frac{\Lambda}{M}\right)^{d_{\rm UV}-1}\left(\frac{M_Z}{\Lambda}\right)^{d-1}\,.
\end{equation}
Using the spectral density $\rho(m^2)=A_d (m^2)^{d-2}$ \cite{Georgi:2007ek}, the propagator is
\cite{Georgi:2007si,Cheung:2007zz,Cheung:2007ap}
\begin{equation} \label{propag}
\left[\Delta_F(q^2)\right]_{\mu\nu}=\frac{A_d}{2\sin(d\pi)}(-q^2)^{d-2}\left(-g_{\mu\nu}+a\frac{q_{\mu}q_{\nu}}{q^2}\right)\,.
\end{equation}
Here $(-q^2)^{d-2}$ is defined as $|q^2|^{d-2}$ for negative $q^2$ and $|q^2|^{d-2}e^{-id\pi}$ for positive $q^2$.
$A_d$ is chosen following the convention of Ref. \cite{Georgi:2007ek}:
\begin{equation}
A_d=\frac{16\pi^{5/2}\Gamma(d+1/2)}{(2\pi)^{2d}\Gamma(d-1)\Gamma(2d)}\,.
\end{equation}
The constant $a=1$ if $O^{\mu}$ is assumed to be transverse, and $a=2(d-2)/(d-1)$ in conformal
field theories \cite{Grinstein:2008qk}. The value of $a$ does not affect the results of this paper.

It should be noted that operators of a conformal field theory are subject to
lower bounds on their scaling dimensions from unitarity, and in particular $d\ge3$
for vector operators \cite{Grinstein:2008qk,Mack:1975je}. However, this bound can be
violated for a hypothetical scale invariant field theory that is
not conformally invariant (see e.g. \ocite{Nakayama:2007qu}). We focus on the range 
$1<d<2$ since unparticle effects are relatively suppressed for higher values of $d$.
(Also, SM contact interactions induced by messenger exchange at the scale $M$
generally dominate over unparticle interference effects for $d\ge3$ \cite{Grinstein:2008qk}.)

A bound on the scale of $O^{\mu}$ interactions can be obtained
from mono-photon production ($e^+e^-\to\gamma+{\rm unparticle}$) at LEP2.
The cross-section is given by \cite{Cheung:2007zz}
\begin{equation} \label{monopho}
\ud\sigma=\frac{A_d e^2 c^2}{8\pi^3 M_Z^2 E_\gamma s}\left(\frac{s-2\sqrt{s}E_\gamma}{M^2_Z}\right)^{d-2}
\frac{s-2\sqrt{s}E_\gamma+(1+\cos^2\theta_\gamma)E^2_\gamma}{1-\cos^2\theta_\gamma}\ud E_\gamma \ud\Omega\,,
\end{equation}
where $c\equiv\sqrt{c^2_V+c^2_A}$, $E_\gamma$ is the photon energy, and $\theta_\gamma$ is the polar angle.
Following \ocite{Cheung:2007ap}, we obtain an upper bound on $c$ using the 
L3 95\% C.L. upper limit $\sigma\simeq0.2$ pb 
(obtained under the cuts $E_\gamma>5$ GeV and $|\cos\theta_\gamma|<0.97$ at $\sqrt{s}=207$ GeV) \cite{Achard:2003tx}.
This ``mono-photon bound'' corresponds to $c<0.026$, 0.032 and 0.057 for $d=1.1$, 1.5 and 1.9 
respectively. Note that since $\Lambda<M$ and unparticle effects can only be probed if 
$\sqrt s<\Lambda$, $c\gtrsim(M_Z/\sqrt s)^{d-1}$ is theoretically inaccessible. This implies
that the current bound from mono-photon production is only relevant for $d\lesssim2.6$ 
(see \fig{boundm}).\footnote{Mono-$Z$ 
production is also considered in Refs. \cite{Cheung:2007ap,Chen:2007qr}. Similarly to mono-photon production,
upper bounds on $c$ can be obtained using the L3 limit on $Z+\,$missing energy cross-section \cite{Acciarri:1999bz},
but they are weaker than the mono-photon bounds.}

Another process considered in \ocite{Cheung:2007zz} is $Z\to q\bar{q}+{\rm unparticle}$,
which contributes to the $Z$ hadronic width.
Here we note that it is important to consider the vertex correction together with
the real emission process, since the two contributions largely cancel each other for 
values of $d$ close to 1 and the former contribution dominates for 
values of $d$ close to 2. As explained in Appendix \ref{hadronic}, the constraint on
unparticles from the $Z$ hadronic width is also weaker than the mono-photon bound.

\EPSFIGURE[t]{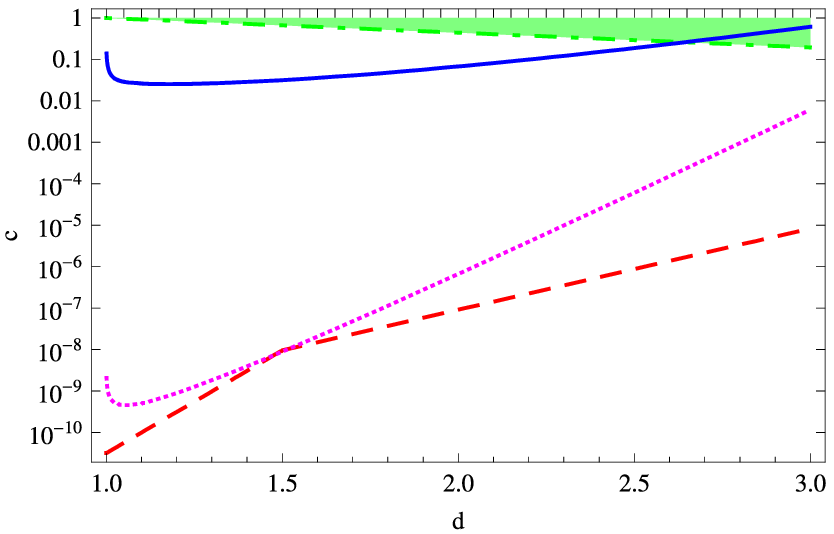, width=11cm}{\label{boundm} Upper bounds on $c$ from mono-photon production (solid blue curve),
BBN (dashed red curve) and SN 1987A (dotted magenta curve). The shaded region corresponds to the theoretically forbidden region $c>(M_Z/\sqrt s)^{d-1}$.}

We now compare the mono-photon bound with the constraints on vector unparticles 
from cosmology and astrophysics \cite{Davoudiasl:2007jr}.
To preserve the successful predictions of Big Bang Nucleosynthesis (BBN),
we require the unparticle sector to be colder than SM radiation
during BBN, so that its energy density is subdominant. For the operator in \eq{veceff}, 
the interaction rate $\Gamma_{\psi}$ redshifts more slowly than the Hubble parameter ${\cal H}$ if $d\le3/2$. 
The unparticle sector can then remain cold if it is decoupled throughout BBN,
corresponding to $\Gamma_{\psi}\lesssim {\cal H}$ for $T\sim1$ MeV.
For $d>3/2$, $\Gamma_{\psi}$ redshifts faster than ${\cal H}$. In this case we require the unparticle
sector to decouple before $T\sim1$ GeV so that the QCD phase-transition only heats up
SM radiation \cite{Davoudiasl:2007jr}.
The BBN constraint, corresponding to
\begin{equation}
\left(\frac{c}{M_Z^{d-1}}\right)^2\lesssim
\frac{T^{3-2d}}{10^{18}{\rm~GeV}}\,\left\{\begin{array}{l}
T\sim1\rm{~MeV~if~}d\le3/2 \\T\sim1\rm{~GeV~if~}d>3/2\end{array}\,,\right. 
\end{equation}
is much more stringent than the mono-photon bound (see \fig{boundm}).
The SN 1987A constraint on unparticle emission 
\cite{Davoudiasl:2007jr,Freitas:2007ip,Hannestad:2007ys,Das:2007nu},
\begin{equation}\label{sn1}
C_d\left(\frac{c}{M_Z^{d-1}}\right)^2\lesssim
4\cdot10^{-22}T_{SN}^{2-2d}\,,
\end{equation}
where the supernova core temperature is taken to be $T_{SN}=30$ MeV
and $C_d\simeq0.01$ (see Appendix \ref{sna}), is similar in magnitude to the BBN constraint.

\section{Broken scale invariance} \label{break}
The BBN and SN 1987A constraints can be evaded provided 
scale invariance is broken at a scale $\mu$ sufficiently large compared to the
relevant energy scales ($\simeq1$ MeV and $\simeq T_{SN}$ respectively).\footnote{Although 
unparticles are stable if scale invariance is exact,
it is not clear if they remain so when scale invariance is broken.
If they are stable, and if $\mu$ is less than the top quark mass, 
it is not sufficient that they decouple at $\sim1$ GeV for $d>3/2$.
Instead, they should remain out of equilibrium at all temperatures before BBN, at least up to
the reheating temperature \cite{McDonald:2007bt}.} We can model broken
scale invariance by removing modes with energy less than $\mu$ in the spectral density,
so that \cite{Fox:2007sy}
\begin{eqnarray} \label{spectral}
\rho(m^2)&=&A_d \theta(m^2-\mu^2)(m^2-\mu^2)^{d-2}\,,\\ \label{propag2}
\left[\Delta_F(q^2)\right]_{\mu\nu}&=&\frac{A_d}{2\sin(d\pi)}[-(q^2-\mu^2)]^{d-2}\left(-g_{\mu\nu}+a\frac{q_{\mu}q_{\nu}}{q^2}\right)\,,
\end{eqnarray}
where $[-(q^2-\mu^2)]^{d-2}$ is defined as $|q^2-\mu^2|^{d-2}$ for $q^2<\mu^2$ and $|q^2-\mu^2|^{d-2}e^{-id\pi}$ for $q^2>\mu^2$.

Due to Boltzmann suppression of the emission, the SN 1987A constraint
with scale invariance broken at a scale $\mu$
corresponds to replacing $C_d$ in \eq{sn1} by (see Appendix \ref{sna}):
\begin{equation} \label{lowc2}
C_d\approx\frac{A_d}{2^{(9/2)-d}\pi^{7/2}(d-1)^{2-d}}
\left(\frac{\mu}{T_{SN}}\right)^{d+5/2}e^{-\mu/T_{SN}}\,.
\end{equation}
Assuming $c$ is close to the mono-photon bound,
the SN 1987A constraint can be evaded provided $\mu\gtrsim1$ GeV.
Note that  other constraints 
arising from long range forces \cite{Deshpande:2007mf}, contributions to the muon and electron anomalous
magnetic moments \cite{Cheung:2007zz,Liao:2007bx}, modifications to positronium decay \cite{Liao:2007bx},
neutrino decay into unparticles \cite{Anchordoqui:2007dp}, and contributions 
to low energy neutrino-electron scattering amplitudes \cite{Balantekin:2007eg} 
are also evaded in this case. 

The mono-photon bound is also modified when scale invariance is broken: In \eq{monopho},
the numerator inside the parentheses is replaced by $s-2\sqrt{s}E_\gamma-\mu^2$,
and the end-point for $E_\gamma$ is shifted from $\sqrt{s}/2$ to $(s-\mu^2)/2\sqrt{s}$.
However, these modifications do not change the cross-section appreciably for
$\mu<M_Z$.

Whether scale invariance is broken or not is relevant for the allowed range of the 
vector unparticle scale dimension $d$.
Consider the decay width from the interaction Eq.\,(\ref{veceff}) where an initial fermion with
mass $m_f$ decays into a massless fermion and the unparticle. Following
\ocite{Choudhury:2007js}, we obtain
\begin{equation}
\frac{\ud\Gamma}{\ud E}=\frac{A_dc^2}{4\pi^2M_Z^{2d-2}}
\frac{E^2[(2+a)m^2_f-4m_fE]}{(m_f^2-2m_fE)^{3-d}}\theta(m_f-2E)\,,
\end{equation}
where $E$ is the energy of the final fermion. Integrating over ${\rm d}E$, it
follows that the total decay width diverges for $d<2$. This is due to the extra
$(1/q^2)$ factor associated with the vector propagator.  However, once
scale invariance is broken, values of $q^2<\mu^2$ are removed from the phase space:
\begin{equation}
\frac{\ud\Gamma}{\ud E}=\frac{A_dc^2}{4\pi^2M_Z^{2d-2}}
\frac{E^2[(2+a)m^2_f-4m_fE]}{(m_f^2-2m_fE)(m_f^2-2m_fE-\mu^2)^{2-d}}\theta(m_f^2-2m_fE-\mu^2)\,,
\end{equation}
and the total width remains finite and positive for $d<2$.\footnote{For 
scalar unparticles and $d<1$, the divergence pointed out in \ocite{Georgi:2007ek}
remains whether $\mu=0$ or not.} 

Next, we discuss how scale invariance could be broken such that $1{\rm
~GeV}\lesssim\mu<M_Z$, first considering the influence of scalar unparticles
and then assuming only vector unparticles couple to the SM. 

\subsection{Scalar unparticles} \label{scalaru}
As pointed out in Ref. \cite{Fox:2007sy}, scale invariance is broken by the operator
\begin{equation} \label{c2}
C_2\frac{\Lambda^{d_{UV}-d}}{M^{d_{UV}-2}}H^{\dag}H O\,,
\end{equation}
where $H$ is the SM Higgs doublet, at an energy scale 
\begin{equation} \label{muu}
\mu\simeq\left[C_2v^2\left(\frac{\Lambda}{M}\right)^{d_{UV}-2}\Lambda^{2-d}\right]^{1/(4-d)}\,,
\end{equation}
where $v=174$ GeV.
Having an experimentally accessible conformal window $\mu\ll\Lambda\sim v$ requires
$C_2\ll1$ \cite{Bander:2007nd}. Assuming $\mu<M_Z$, another upper bound on 
$\mu$ and $C_2$ can be obtained from the threshold correction to
the fine structure constant \cite{Bander:2007nd}. If the operator 
\begin{equation} \label{c4}
C_4\frac{\Lambda^{d_{UV}-d}}{M^{d_{UV}}}F^{\rho\delta}F_{\rho\delta} O
\end{equation}
exists, the value of $\alpha^{-1}(M_Z)$ remains within the current uncertainty for
\begin{equation} \label{upbound}
\mu\lesssim\left(\frac{M}{\Lambda}\right)^{d_{UV}/d}\left(\frac{1}{10^{4.5}C_4}\right)^{1/d}\Lambda\,.
\end{equation}

Eq. (\ref{upbound}) provides an upper bound on $\mu$, whereas Eqs. (\ref{sn1},
\ref{lowc2}) provide a lower bound. There can be a scale invariant window below
$M_Z$ between these two bounds without violating any of the other constraints discussed above.  
As a specific example we take $d_{UV}=2$ or 3, $C_V=C_A=1$ and
$\Lambda=v$.  Setting $M$ equal to the mono-photon bound using Eqs.
(\ref{convert}, \ref{monopho}), we calculate the range of $C_2$, $C_4$ and $\mu$
that satisfies the other constraints. As shown in Table \ref{tablo} and
\fig{c2c4}, there is an allowed range of $\mu$ below $M_Z$, provided the
scalar unparticle operators couple somewhat weakly ($C_4,C_2\lesssim0.1$)
compared to vector operators ($C_V=C_A=1$).

\TABLE[t]{
\resizebox{!}{1.4cm}{
\begin{tabular}
{r@{\hspace{.7cm}}r@{\hspace{.7cm}}r@{\hspace{.7cm}}r@{\hspace{.7cm}}r@{\hspace{.7cm}}r@{\hspace{.7cm}}r}
&\multicolumn{3}{c}{$d_{UV}=2$}&\multicolumn{3}{c}{$d_{UV}=3$} \\
\hline
$d$ & $M$ (GeV) & $\mu$ (GeV) & $C_2$ & $M$ (GeV) & $\mu$ (GeV) & $C_2$ \\
\hline\hline
1.1 & 8800 & 1.5--$M_Z$ & $9\times10^{-7}$--$0.15$  & 1200 & 1.5--35 & $7\times10^{-6}$--$0.07$  \\
\hline
1.5 & 5600 & 1.3--74 & $4\times10^{-6}$--$0.12$   & 990 & 1.3--31 & $2\times10^{-5}$--$0.08$ \\
\hline
1.9 & 2400 & 1.1--49 & $2\times10^{-5}$--$0.07$   & 650 & 1.1--25 & $8\times10^{-5}$--$0.06$ \\
\hline
\end{tabular}
}
\caption{\label{tablo}
The allowed range of $\mu$ (below $M_Z$) and $C_2$ for $M$ at the 
mono-photon bound, assuming $C_V=C_A=1$ (see \eq{vecop}), $\Lambda=v$ and $C_4=C_2$ (see Eqs. 
(\ref{c2}, \ref{c4})).}
}

\EPSFIGURE[t] 
{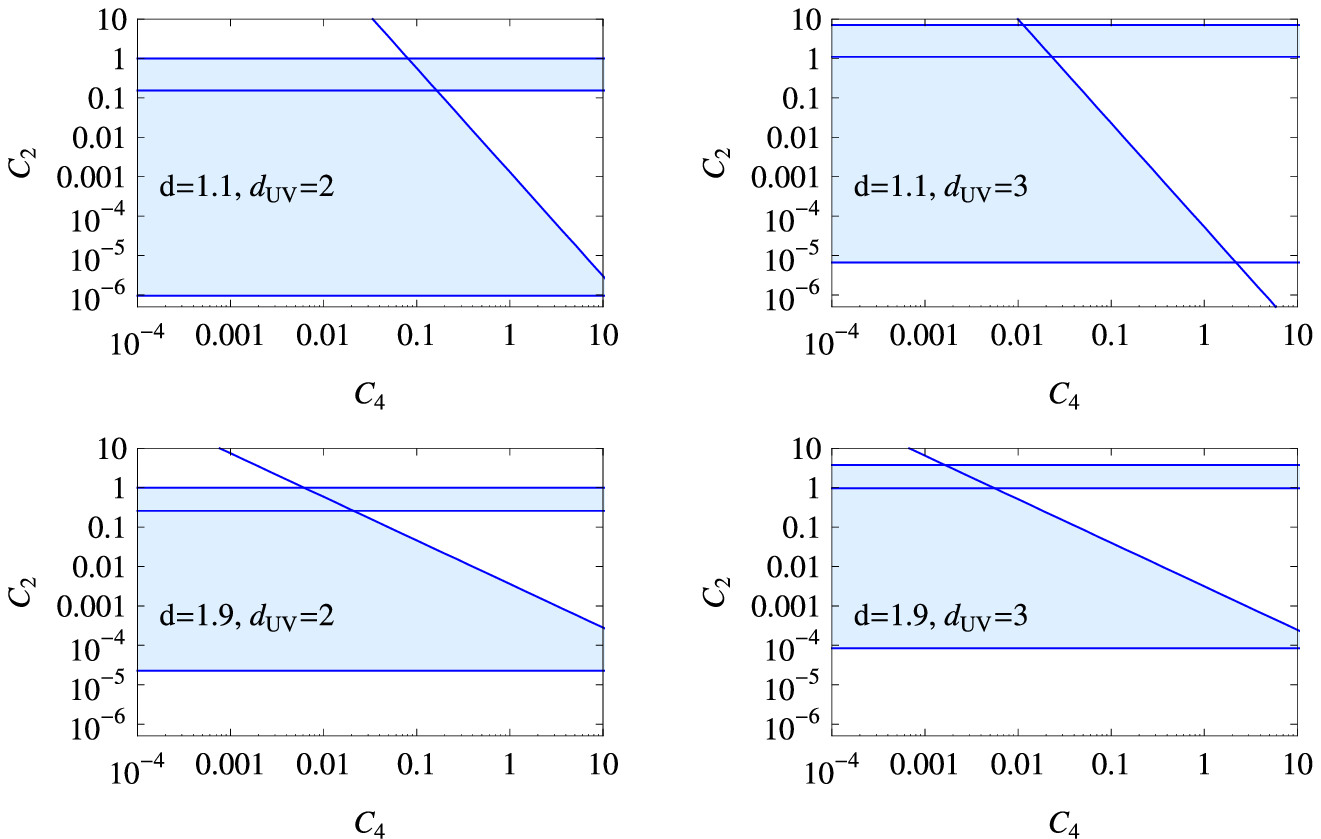, width=15cm} 
{\label{c2c4} The shaded region in the $C_4$--$C_2$ plane is allowed for 
$M$ at the  mono-photon bound ($C_V=C_A=1,\;\Lambda=v$). 
The lower horizontal line corresponds to the SN 1987A constraint, the middle
horizontal line corresponds to $\mu=M_Z$, and the upper line to
 $\mu=v$. The diagonal line is the constraint from $\alpha(M_Z)$.}

\subsection{Vector unparticles} \label{vectoru}
Even if only vector unparticles exist, scale invariance can still be broken
if the Higgs couples to higher-dimensional operators such as $O^{\mu}O_{\mu}$.
Furthermore, due to the higher dimensionality, the scale $\mu$ is naturally
suppressed compared to the electroweak scale.\footnote{See
\ocite{Zwicky:2007vv} for a similar scenario with charged scalar unparticles.}
 Consider the operators
\begin{equation} \label{vecop2}
\frac{\Lambda^{2d_{UV}-d_*}}{M^{2d_{UV}-2}}H^{\dag}HO^{\mu}O_{\mu}
+\frac{\Lambda^{2d_{UV}-d_*}}{M^{2d_{UV}}}F^{\rho\delta}F_{\rho\delta}O^{\mu}O_{\mu}
\end{equation}
where we have set $C_2=C_4=1$, and the scale dimension of $O^{\mu}O_{\mu}\equiv d_*\le2d$.
\eq{muu} and \eq{upbound} are modified as follows:
\begin{eqnarray} \label{muu2}
\mu&\simeq&\left[\left(\frac{\Lambda}{M}\right)^{2d_{UV}-2}\Lambda^{2-d_*}v^2\right]^{1/(4-d_*)}\,,
\\ \label{upbound2}
\mu&\lesssim&\left(\frac{M}{\Lambda}\right)^{2d_{UV}/d_*}10^{-4.5/d_*}\Lambda\,.
\end{eqnarray}
As shown in \fig{dstar}, $\mu$ can easily lie in the allowed range.

\EPSFIGURE[t] 
{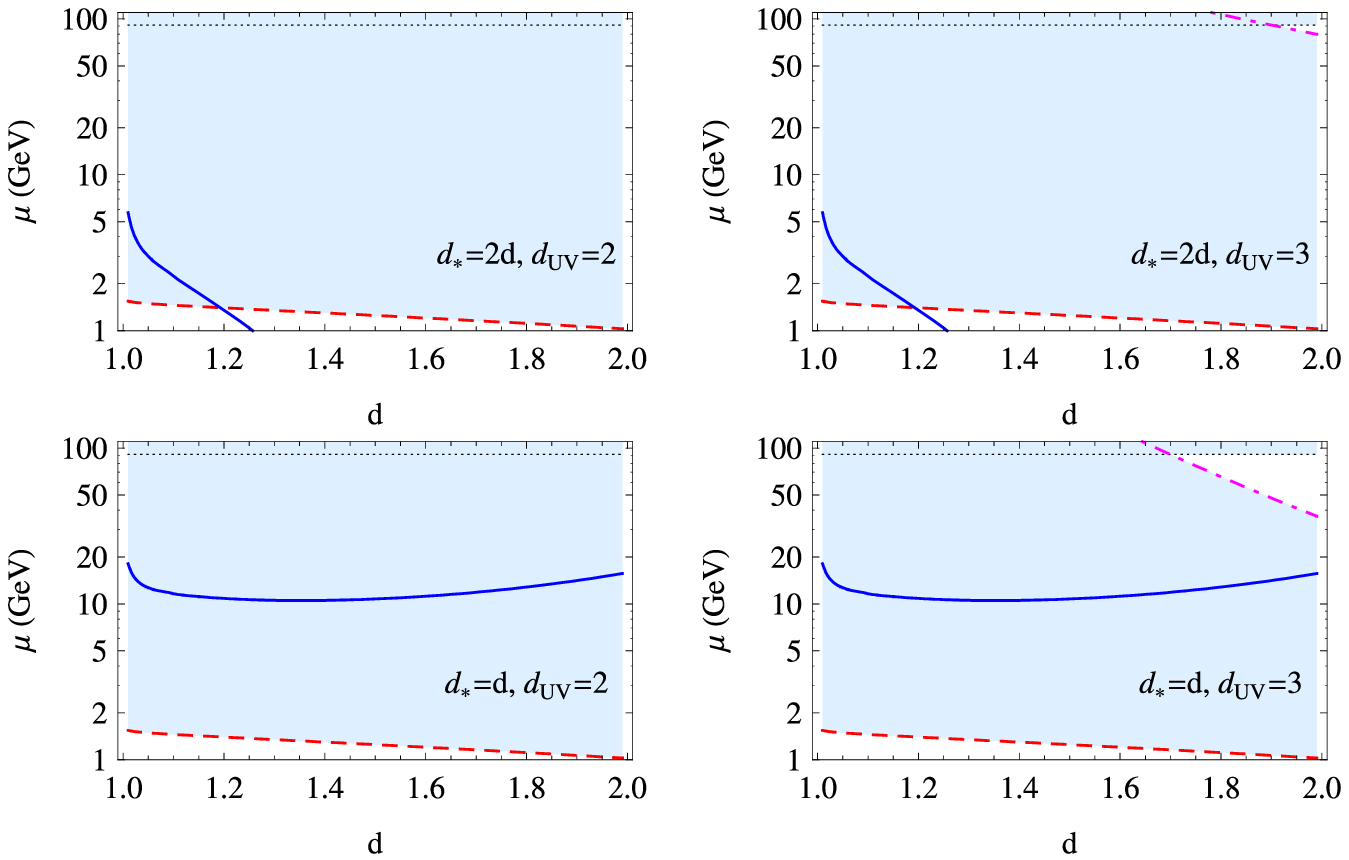, width=15cm} 
{\label{dstar} The shaded regions in the $d$--$\mu$ plane are allowed for 
$M$ at the mono-photon bound ($\Lambda=v$, $C_V=C_A=1$). The solid blue curve is $\mu$
as given by \eq{muu2}.
The dashed red curve corresponds to the SN 1987A constraint, the dotted-dashed
magenta curve corresponds to the constraint from $\alpha(M_Z)$, and the horizontal
dotted line corresponds to $\mu=M_Z$.}

\subsection{Implications for Higgs phenomenology} \label{higgs}
The effects of scalar unparticles on Higgs phenomenology
have been considered in Refs. \cite{Kikuchi:2007qd,Delgado:2007dx}.\footnote{See
also \ocite{Deshpande:2007jy} for supersymmetric unparticle effects.}
For scalar unparticles the same operator $H^{\dag}HO$ is responsible
for breaking scale invariance and Higgs-unparticle mixing to lowest
order, and thus the effects are suppressed for $\mu\ll M_Z$.
To be more explicit, the mixing between the SM Higgs boson $h$ and the unparticle
is induced by the interaction term $(\mu^{4-d}/v)Oh$. Considering
the effective Higgs coupling $(1/v)C_{\gamma\gamma}hF_{\mu\nu}F^{\mu\nu}$
as an example, the contribution from the above interaction and \eq{c4} is
given by \cite{Kikuchi:2007qd} 
\begin{equation}
C_{\gamma\gamma}(h\to O\to\gamma\gamma)\simeq C_4 \frac{e^{-id\pi}A_d}{2\sin d\pi} 
\left(\frac{\mu}{m_h}\right)^{4-d}
\left(\frac{m_h}{\Lambda}\right)^{d}
\left(\frac{\Lambda}{M}\right)^{d_{UV}}\,.
\end{equation}
Provided $\mu\ll M_Z$, this is small compared to the SM 
effective coupling $C_{\gamma\gamma}\sim10^{-3}$. The Higgs partial decay
width to fermions induced by the operator $\bar{\psi}\gamma_{\mu}
D^{\mu}\psi O$ is similarly suppressed.

On the other hand, for vector unparticles scale invariance is broken
by $H^{\dag}HO^{\mu}O_{\mu}$ whereas mixing is (also) induced by
\begin{equation} \label{vecop3}
\frac{\Lambda^{d_{UV}-d}}{M^{d_{UV}-1}}H^{\dag}D_{\mu}HO^{\mu}\,.
\end{equation}
Using \eq{vecop} and \eq{vecop3}, the effective fermionic operator is 
\begin{equation}
\frac{1}{\Lambda^2_{\rm eff}}H^{\dag}D_{\mu}H\bar{\psi}\gamma^{\mu}\psi\,,
\end{equation}
where
\begin{equation}
\frac{1}{\Lambda^2_{\rm eff}}=\frac{1}{s}\frac{e^{-id\pi}A_d}{2\sin d\pi}\left(\frac{\sqrt{s}}{\Lambda}\right)^{2d-2}
\left(\frac{\Lambda}{M}\right)^{2(d_{UV}-1)}\,.
\end{equation}
Contributions of this operator to Higgs production at a linear collider have been considered in \ocite{Kile:2007ts}.
The main effect is interference with the SM Higgs-strahlung (HZ) cross-section,
which can be substantial in the $e^+e^-\to h\mu^+\mu^-$, $h\tau^+\tau^-$, $h\bar{q}q$
channels for $M$ close to the mono-photon bound.

It is also interesting to note that the operator $H^{\dag}HO^{\mu}O_{\mu}$
induces a partial decay width $\Gamma(h\to O^{\mu}O_{\mu})\sim\mu^{8-2d_*}/(v^2 m_h^{5-2d_*})$.
Although this is typically small, it becomes of order $m^3_h/v^2$ in the limit $d_*\to4$
(i.e. $d_*\to2d$ and $d\to2$). With decays of $O^{\mu}O_{\mu}$ suppressed, the
Higgs would then decay invisibly.

\section{Muon pair production bounds on vector unparticles} \label{collide}
We have already obtained a collider bound using \eq{monopho}.
Other bounds can be obtained using the ratio 
$R_U\equiv\sigma{\rm(with~unparticles)}/\sigma{\rm(without~unparticles)}$
as well as the FBA (defined in Appendix \ref{isqed}) 
for $e^+e^-\to\mu^+\mu^-$, and by combining measurements at and away from the Z pole.
As shown in \ocite{Georgi:2007si} and discussed further in \ocite{Luo:2007me},
vector couplings of unparticles will mainly affect $R_U$ away from the Z pole,
and FBA at the Z pole. Axial-vector couplings
have the opposite behaviour.

Due to the resonance-like behavior 
at $\mu$ (referred to as ``un-resonance'' \cite{Rizzo:2007xr}), measurements at energies around $\mu$
would be particularly sensitive to unparticle effects. Thus the bounds on $c_{V,A}$ 
(defined in Eq. (\ref{veceff})) for a
given value of $d$ will also depend on $\mu$. As an example we plot FBA and $R_U$ for $d=1.1$ in \fig{cvca}.
Taking $c_A=0.026$ and $c_V=0$, FBA$=-7.2\%$ for $\sqrt{s}=34.8$ GeV if $\mu=30$ GeV, to be compared
with $-8.3\%$ if $\mu\ll30$ GeV, and $-8.9\%$ for SM. Taking into account the measurement  
FBA$ = -10.4 \pm1.3 \pm0.5 \%$ at the same center-of-mass
energy \cite{dataref1}, it is clear that the bound on $c_A$ for $\mu=30$ GeV will be
more stringent compared to the bound for $\mu\ll30$ GeV (see \fig{vcp}).

\EPSFIGURE[t] 
{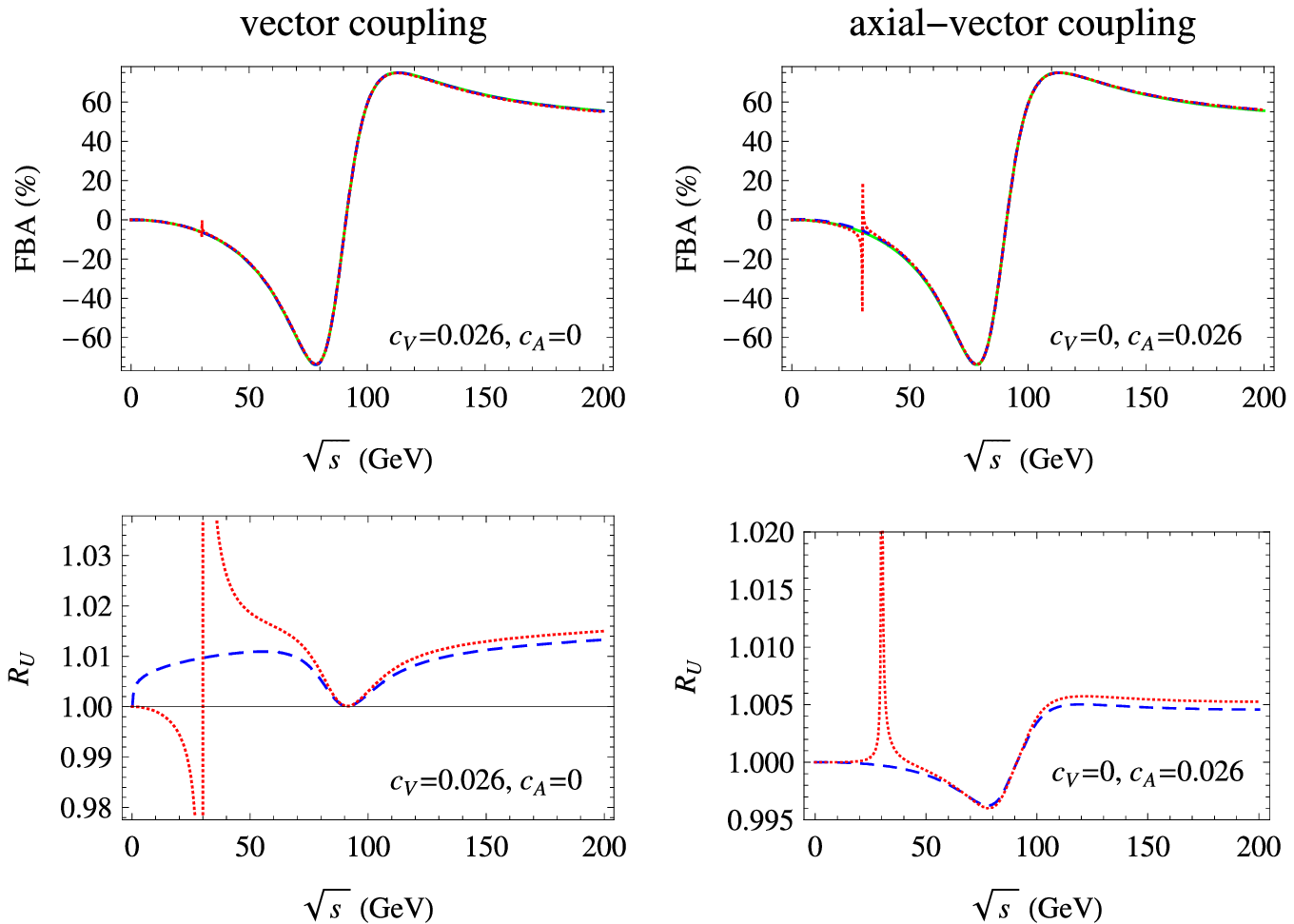, width=15cm} 
{\label{cvca}
FBA and $R_U$ for $e^+e^-\to\mu^+\mu^-$ with $d=1.1$. 
Solid green curves: SM; dashed blue curves: unparticles with
$\mu=0$, dotted red curves: unparticles with $\mu=30$ GeV. ($c_{V,A}=0.026$ correspond
to the mono-photon bound of section \ref{bound}.)} 

It should be noted that for the propagator in \eq{propag2}, the area under the un-resonance
diverges for $d<1.5$. However, it is likely that once scale invariance is broken, particle-like
modes will appear in the spectral density \cite{Fox:2007sy}. For example, vacuum polarization
correction from fermion loops will modify \eq{propag2} as follows:
\begin{equation} \label{vaceq}
\frac{1}{(q^2-\mu^2)^{2-d}}\to\frac{1}{(q^2-\mu^2)^{2-d}-\Pi(q^2)}
\end{equation}
It can therefore be expected that the unparticle will become unstable, and the
area under the un-resonance will depend on the decay width.

We have performed a $\chi^2$ analysis of LEP1-Aleph, KEK-Venus and PETRA-MarkJ 
$e^+e^-\to\mu^+\mu^-$ cross-section and FBA data \cite{dataref1,datarefs}.
The simulation includes the vacuum polarization correction from fermion loops to the
unparticle propagator (see Appendix \ref{vacpol}) and uses a fixed Z decay
width $\Gamma_Z=2.41$ GeV which is the SM best-fit value for the data.
Initial-state QED corrections are also included
(see Appendix \ref{isqed}).

The allowed regions in the $c_V$--$c_A$ plane for different values of $d$ and
$\mu$ are shown in \fig{vcp}. The best-fit parameters and $\chi^2$ values are
listed in Table \ref{sort}, and fits to FBA data with and without unparticles
are displayed in \fig{fba}. For values of $d$ close to 1 where fermion-unparticle couplings are
less suppressed by $M_Z^{1-d}$, constraints on $c_V$ and $c_A$ are more
stringent and the dependence on $\mu$ is more significant. 
The mono-photon bound discussed
in \sektion{bound} is stronger than the muon pair production bound for 
$d\gtrsim1.3$.\footnote{Recently, it was noted that processes mediated by unparticle self-interactions
lead to multi-body final states which could be the most promising modes for
unparticle discovery at colliders \cite{strasslerfeng}.
However, details of the hidden sector are required to make predictions.}

Finally, it is worth emphasizing that the spin and scale
dimension of the exchanged unparticle can be probed by analyzing the scattering angle and
energy distributions of differential cross-sections in a linear collider, for
both real emission and virtual exchange processes
\cite{Georgi:2007si,Cheung:2007ap,Chen:2007qr}. Furthermore, for polarized beams the
azimuthal dependence of the final state fermion can provide an independent
measure of the scaling dimension for spin-1 unparticle exchange
\cite{Huitu:2007im}.

\clearpage
\EPSFIGURE[t] 
{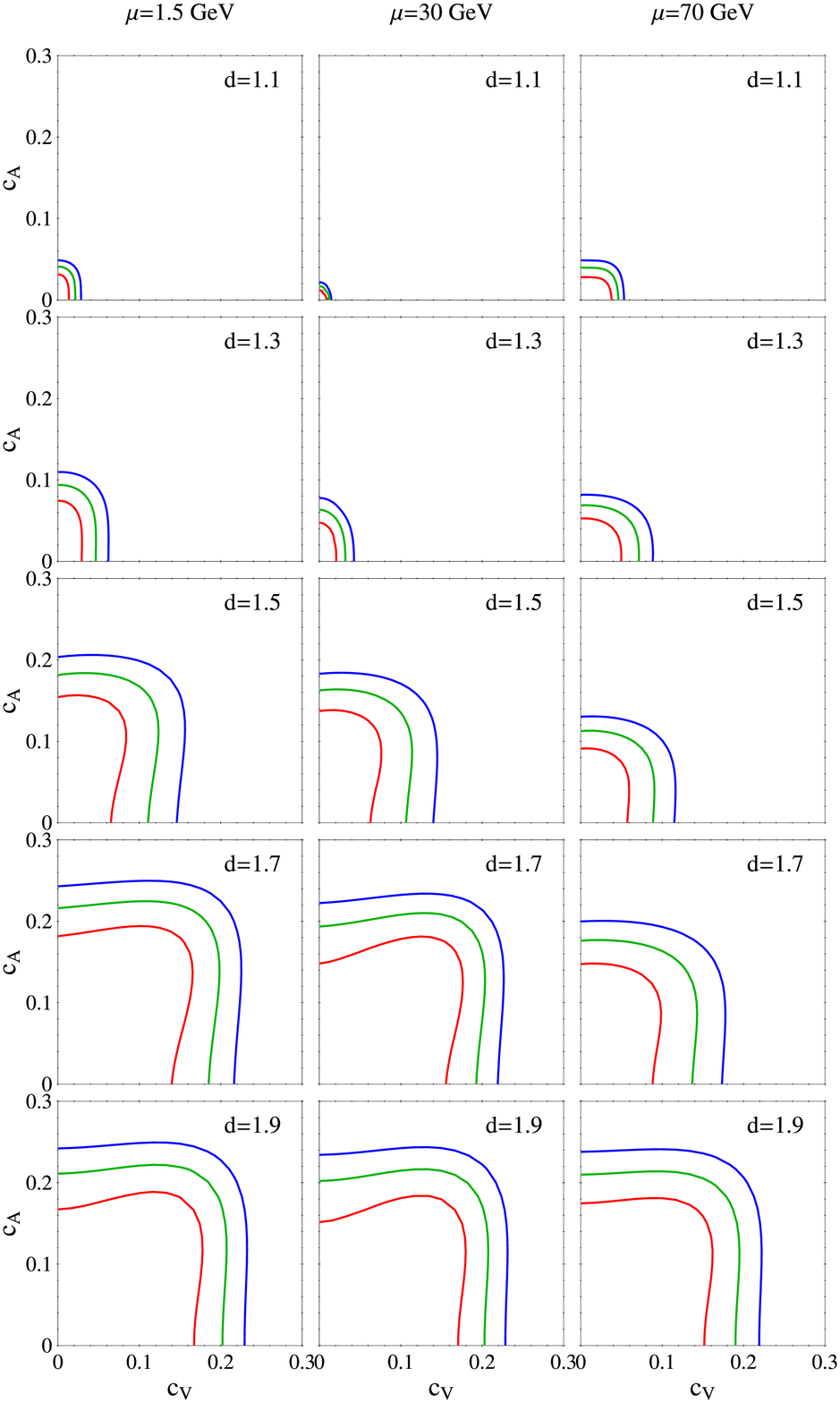, width=14cm,height=21cm} 
{\label{vcp} Allowed regions in the $c_V$--$c_A$ plane from a $\chi^2$
analysis of $e^+e^-\to\mu^+\mu^-$ cross-section and FBA data. The contours represent
the 1$\sigma$, 2$\sigma$ and 3$\sigma$ regions. We only show results in the first
quadrant since the dependence of the cross-section and
FBA on the relative sign of $c_V$ and $c_A$ is too weak to be visible.}
\clearpage

\TABLE[t]{
\resizebox{!}{1.97cm}{
\begin{tabular}
{r@{\hspace{.7cm}}r@{\hspace{.7cm}}r@{\hspace{.7cm}}r@{\hspace{.7cm}}r@{\hspace{.7cm}}r@{\hspace{.7cm}}r@{\hspace{.7cm}}r@{\hspace{.7cm}}r@{\hspace{.7cm}}r}
&\multicolumn{3}{c}{$\mu=1.5$ GeV}&\multicolumn{3}{c}{$\mu=30$ GeV}&\multicolumn{3}{c}{$\mu=70$ GeV} \\
\hline
$d$ & $c_V$ & $c_A$ & $\chi^2$ & $c_V$ & $c_A$ & $\chi^2$ & $c_V$ & $c_A$ & $\chi^2$\\
\hline\hline
SM & -- & -- & 154.3 & -- & -- & 154.3& -- & -- & 154.3 \\
\hline
1.1 &0.001&$5\times10^{-4}$&154.3&$2\times10^{-4}$&0.002&154.3&0.018&$4\times10^{-4}$&154.0\\
\hline
1.3 &0.003&0.02&154.2&$1\times10^{-4}$&0.02&154.2&0.0081&0.020&154.3\\
\hline
1.5 &0.0089&0.093&153.6&0.0059&0.081&153.6&0.0016&0.036&154.2\\
\hline
1.7&0.083&0.13&153.1&0.12&0.12&152.0&0.0071&0.081&153.9\\
\hline
1.9&0.11&0.11&152.6&0.12&0.11&152.1&0.085&0.11&153.4\\
\hline
\end{tabular}
}
\caption{\label{sort} Best-fit parameters and $\chi^2$ values from
an analysis of LEP1-Aleph, KEK-Venus and PETRA-MarkJ
$e^+e^-\to\mu^+\mu^-$ cross-section and FBA data.
The dataset is comprised of 55 FBA data points and 54 cross-section data points.
The $\chi^2$ value for SM is obtained from
a scan on 4 SM parameters $M_Z,\,\Gamma_Z,\,e$ and $\sin\theta_W$. The $\chi^2$ values for
SM + unparticles are obtained from a scan on 2 unparticle parameters $c_V$ and $c_A$ for different
fixed values of $d$ and $\mu$, with SM parameters fixed to their SM best-fit values.
} }

\EPSFIGURE[t] 
{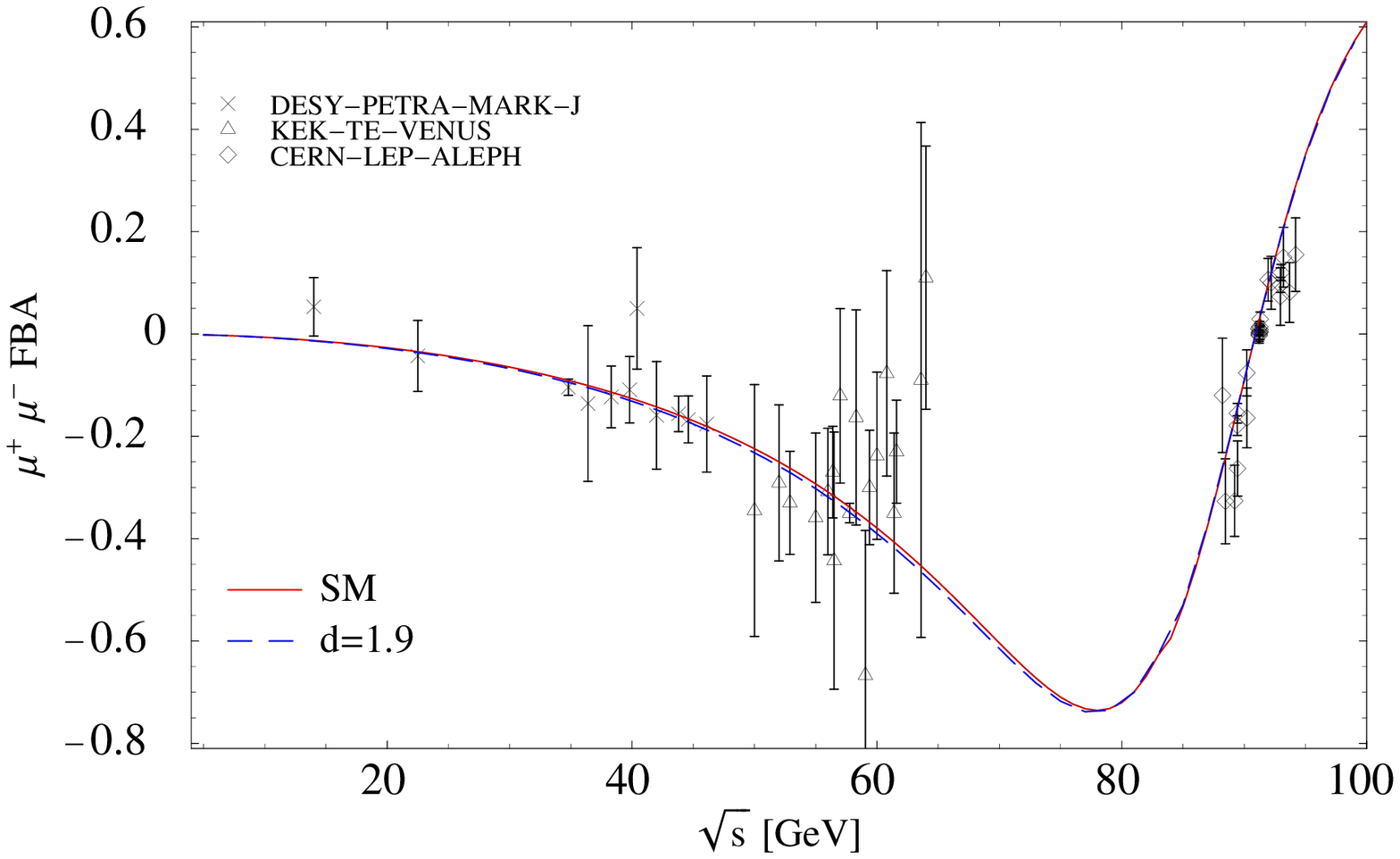, width=14cm} 
{\label{fba} Fits to $e^+e^-\to\mu^+\mu^-$ FBA data,
from a scan over $c_V$ and $c_A$ for different fixed $d$ and $\mu$ values. The solid curve
(red) is the SM fit and the dashed curve (blue) is the best-fit curve for $d=1.9$
and $\mu=1.5$ GeV, $\mu=30$ GeV or $\mu=70$ GeV. 
For $d\lesssim1.5$ the curves with unparticle
contribution are almost indistinguishable from the SM curve. Note that the
unparticle FBA curve does not exhibit divergent behavior at $\mu$ as the vacuum
polarization introduces a finite decay width and stabilizes the unparticle
propagator.} 
\clearpage

\section{Summary}
For exact scale invariance, astrophysical and cosmological constraints
are in gross conflict with the possibility of probing unparticles in colliders.
We showed that for vector unparticles collider constraints 
become relevant only if scale invariance is broken at a scale $\mu\gtrsim1$ GeV.
Breaking the scale invariance also affects collider expectations by giving
rise to a resonance-like behaviour.
On the other hand, unparticle effects cannot be observed at energies below the scale $\mu$. 
We focused on the case $1{\rm~GeV}\lesssim\mu<M_Z$ which allows unparticle
effects to show up in Z exchange observables, and
gave demonstrations of how this can be realized through unparticle--Higgs couplings.

Simple bounds on vector unparticles have been obtained using effective contact interactions
in Refs. \cite{Cheung:2007ap,Bander:2007nd}.
Here we have made a more detailed analysis using $e^+e^-\to\mu^+\mu^-$ cross-section
and forward-backward asymmetry data both at the Z pole and away from it,
also taking into account the resonance-like behaviour associated with broken scale invariance.
We found that unparticle parameters are severely constrained for values of
scale dimension $d$ close to 1. For $d\gtrsim1.3$, constraints
from mono-photon production are more stringent compared to constraints from muon
pair production. 

\acknowledgments
This research was supported by the U.S. Department of Energy (Grant Nos.
DEFG02-95ER40896, DE-FG02-84ER40173 and DE-FG02-04ER41308), by the U.S.
National Science Foundation (Grant No. PHY-0544278), and by the Wisconsin
Alumni Research Foundation.

\appendix
\section{Unparticle contribution to the $Z$ hadronic width}\label{hadronic}
\ocite{Carone:1994aa} studied the real and virtual massive
vector boson contribution to the $Z$ hadronic width $R_Z$. To calculate the constraint on unparticles, 
we write the unparticle operator
in terms of deconstructed particle fields \cite{Stephanov:2007ry}: 
$O^{\mu}=\sum_j F_j \lambda^{\mu}_j$, where the field $\lambda^{\mu}_j$ has 
mass $M_j^2=j\Delta^2$ and 
\begin{equation}
F_j^2=\frac{A_d}{2\pi}\Delta^2(M_j^2)^{d-2}\,.
\end{equation}
In the limit $\Delta\to0$, the contribution to $R_Z$ is obtained by integrating
the contribution from a vector boson with mass $m$ over $\delta=m^2/M^2_Z$:
\begin{equation}
\frac{\Delta R_Z}{R_Z}= \frac{A_dc_V^2}{16\pi^3}
\left[
\int_0^1  \delta^{d-2} F_{1}(\delta) d\delta 
+
\int_0^\infty      \delta^{d-2} F_{2}(\delta) d\delta \right]\ ,
\end{equation}
where \cite{Carone:1994aa}
\begin{eqnarray}
F_{1}(\delta)&=&(1+\delta)^2 \left[3\ln\delta + (\ln\delta)^2\right]
+5(1-\delta^2)-2\delta\ln\delta
\nonumber \\
& &-2(1+\delta)^2
\left[\ln(1+\delta)\ln\delta+\mbox{Li}_2\left(\frac{1}{1+\delta}\right)-
\mbox{Li}_2\left(\frac{\delta}{1+\delta}\right)\right] , \\
F_{2}(\delta)&=&-2\left\{\frac{7}{4}+\delta+(\delta+\frac{3}{2})\ln\delta
\right.
\nonumber \\
& &\left. +
(1+\delta)^2\left[\mbox{Li}_2\left(\frac{\delta}{1+\delta}\right)
+\frac{1}{2}\ln^2\left(\frac{\delta}{1+\delta}\right)
-\frac{\pi^2}{6}\right]\right\} ,
\end{eqnarray}
Li$_2(x) = -\int_0^x \ud t\ln(1-t)/t$ is the Spence
function, and uniform coupling for quarks is assumed. Note that the upper limit 1 of the $\delta$ integration is kinematic
for the real emission, and the upper limit becomes $\infty$ for the
virtual correction.

Evaluating the integrals, we obtain $\Delta R_Z/R_Z\simeq0.01 c_V^2$, 
corresponding to a bound $c_V\lesssim0.3$ since $\Delta R_Z/R_Z=\Delta\alpha_s/\pi\simeq0.001$.
Including the axial-vector coupling is straight-forward and leads to $c\lesssim0.3/\sqrt2$.

\section{The bound from SN 1987A cooling}\label{sna}
As discussed in Refs. \cite{Davoudiasl:2007jr,Freitas:2007ip,Hannestad:2007ys,Das:2007nu,Lewis:2007ss},
SN 1987A energy-loss arguments provide very restrictive constraints on unparticle couplings.
In this section we discuss the constraint from pair annihilation of neutrinos
and obtain the prefactor $C_d$ in Eqs. (\ref{sn1}, \ref{lowc2}) following the method 
in Refs. \cite{Stephanov:2007ry,Lewis:2007ss}.\footnote{The 
constraint from pair annihilation (for exact scale invariance) is discussed in Ref. 
\cite{Hannestad:2007ys}. The constraint from nucleon bremsstrahlung is
similar in magnitude \cite{Davoudiasl:2007jr,Freitas:2007ip,Hannestad:2007ys}.}

The observed duration of SN 1987A neutrino burst puts a constraint on the
supernova volume emissivity \cite{Raffelt:1990yz}
\begin{equation}
Q\lesssim3\times10^{33}{\rm~erg~cm}^{-3}{\rm ~s}^{-1}\,,
\end{equation}
where the supernova core temperature is taken to be $T_{SN}=30$ MeV. This
corresponds to
\begin{equation}
Q\lesssim4\times10^{-22}T_{SN}^5\,.
\end{equation}

As in Appendix \ref{hadronic}, we write the unparticle operator
in terms of deconstructed particle fields. 
The cross-section for neutrino pair annihilation to $\lambda^{\mu}_j$ is
\begin{equation}\label{cros}
\sigma_j=\left(\frac{c}{M_Z^{d-1}}\right)^2A_d\Delta^2(M_j^2)^{d-2}\delta(s-M_j^2)\,.
\end{equation}
The supernova volume emissivity is found by thermally averaging over the
Fermi-Dirac distribution (see e.g. \ocite{Goodman:1986we}):
\begin{equation}\label{emiss}
Q_j=\int\frac{\ud^3\mathbf{k_1}}{(2\pi)^32E_1}\frac{2}{e^{E_1/T}+1}
\int\frac{\ud^3\mathbf{k_2}}{(2\pi)^32E_2}\frac{2}{e^{E_2/T}+1}
(E_1+E_2)2s\sigma_j\,,
\end{equation}
where we ignored chemical potentials (see Ref. \cite{Hannestad:2007ys}),
and $s=2E_1E_2(1-\cos\theta)$. The total emissivity is obtained as\footnote{See Ref.
\cite{Lewis:2007ss} for similar calculations with tensor unparticles.}
\begin{equation}\label{emis2}
Q=\frac{1}{\Delta^2}\int\ud M^2_jQ_j=
C_d\left(\frac{c}{M_Z^{d-1}}\right)^2T_{SN}^{2d+3}\,,
\end{equation}
where
\begin{equation}
C_d=\frac{2^{2d-3}A_d}{\pi^4 d}\int_0^\infty\ud x_1 \ud x_2\frac{(x_1x_2)^d(x_1+x_2)}
{(e^{x_1}+1)(e^{x_2}+1)}\simeq0.01\,.
\end{equation}
We now repeat the calculation for non-zero $\mu$. By matching to the spectral
density in \eq{spectral}, we have
\begin{eqnarray}
F_j^2&=&\frac{A_d}{2\pi}\Delta^2(M_j^2-\mu^2)^{d-2}\theta(M_j^2-\mu^2)\,,\\\label{cros2}
\sigma_j&=&\left(\frac{c}{M_Z^{d-1}}\right)^2A_d\Delta^2
(M_j^2-\mu^2)^{d-2}\theta(M_j^2-\mu^2)\delta(s-M_j^2)\,.
\end{eqnarray}
Using Eqs. (\ref{emiss}, \ref{cros2}), we obtain \eq{emis2} with
\begin{eqnarray}\label{cdd}
C_d&=&\frac{2^{d-3}A_d}{\pi^4}\int_0^\infty\ud x_1\int_{\mu^2/(4T_{SN}^2x_1)}^\infty\ud x_2
\int_{-1}^{1-\mu^2/(2T_{SN}^2x_1x_2)}\ud(\cos\theta)\nonumber\\&&{}
\frac{(x_1x_2)^d(x_1+x_2)(1-\cos\theta)\left(1-\cos\theta-\frac{\mu^2}{2T_{SN}^2x_1x_2}\right)^{d-2}}
{(e^{x_1}+1)(e^{x_2}+1)}\,.
\end{eqnarray}
The approximation \eq{lowc2} is obtained from \eq{cdd} assuming $\mu\gg T_{SN}$.

\section{Vacuum polarization correction}\label{vacpol}
To lowest order, $\Pi(q^2)$ in \eq{vaceq} is given as follows:
\begin{eqnarray}
\Pi&=&\Pi_{LL}+\Pi_{LR}+\Pi_{RL}+\Pi_{RR}\,,\\
\Pi_{LR}&=&\Pi_{RL}=-2\frac{c_Lc_R}{16\pi^2M_Z^{2d-2}}
\int^1_0\ud x\,m_f^2\log\left(\frac{m^2_f}{m^2_f-x(1-x)q^2}\right)\,,\\
\Pi_{LL/RR}&=&-4\frac{c_{L/R}c_{L/R}}{16\pi^2M_Z^{2d-2}}
\int^1_0\ud x\left(x(1-x)q^2-\frac12m^2_f\right)\log\left(\frac{m^2_f}{m^2_f-x(1-x)q^2}\right)\,,
\end{eqnarray}
where $c_{L}=c_V-c_A$, $c_{R}=c_V+c_A$ and $m_f$ is the mass of the
fermion in the loop. $\Pi(q^2)$ is complex for the $s$ channel with $q^2>4m^2_f$,
and the imaginary part will stabilize the propagator when the real part coincides with the pole.
We assume a universal coupling between the unparticle and different fermions that include charged
leptons, neutrinos and quarks. A numerical example for $\Pi(q^2)$ that is summed over 
the fermions is shown in Fig.~\ref{vacvac}.

\EPSFIGURE[t]{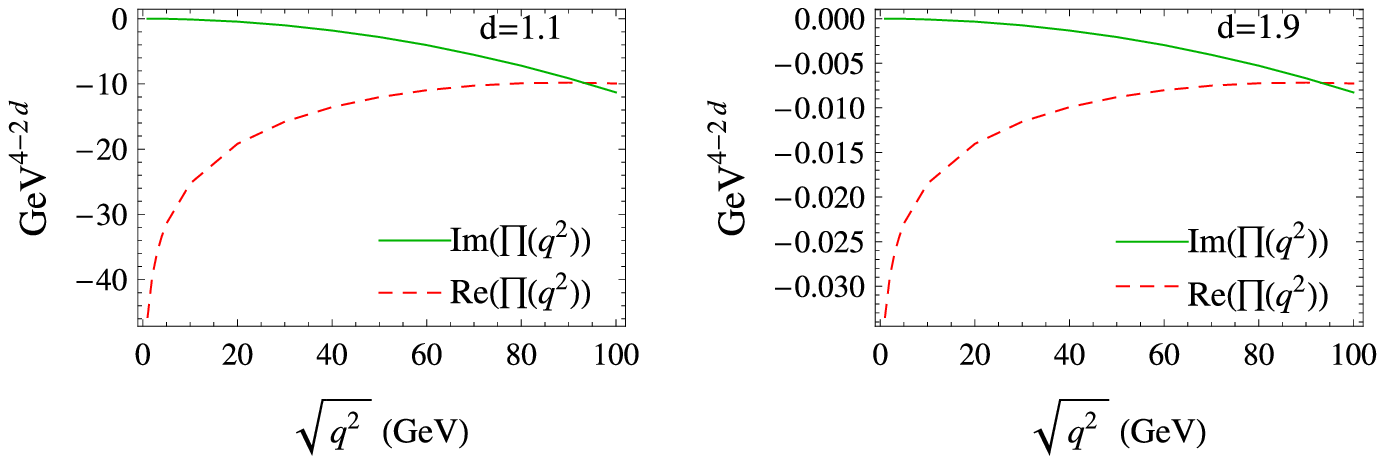, width=15cm}{\label{vacvac} $\Pi(q^2)$ from charged lepton, neutrino 
and quark loops, assuming  fermion couplings $c_V=c_A=0.05$.}

\EPSFIGURE[t] 
{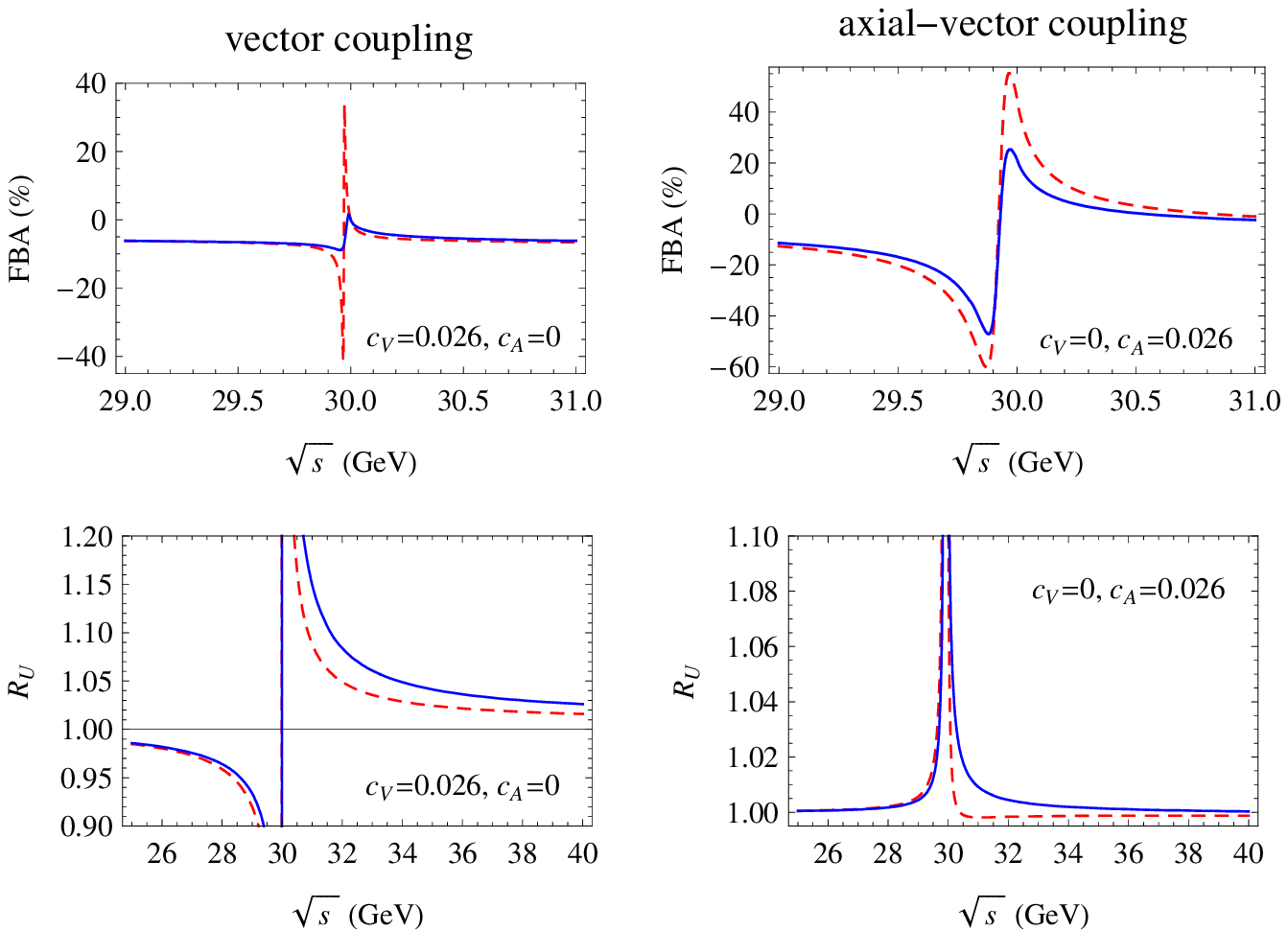, width=15cm} 
{\label{cvcazoom}
FBA and $R_U$ for $e^+e^-\to\mu^+\mu^-$, $d=1.1$ and $\mu=30$ GeV. 
Solid blue curves: with initial state QED corrections to the unparticle exchange term  
as well as the interference terms between $\gamma,Z$ and the unparticle. Red dashed curves:
without initial state QED corrections.
($c_{V,A}=0.026$ correspond
to the mono-photon bound of section \ref{bound}.)} 

\section{Initial state QED corrections}\label{isqed}
Initial state QED corrections significantly affect the cross-section and FBA around $\mu$
(see \fig{cvcazoom}). Since the corrections to the SM cross-section $\sigma_{SM}$ are removed from 
the KEK-Venus and PETRA-MarkJ data,
we only consider the corrections to the unparticle exchange term $\sigma_U$ and the interference 
terms $\sigma_{\rm int}$ between $\gamma,Z$ and the unparticle. The corrected cross section is obtained
by convoluting the relevant terms with a radiator function $H(x)$:
\begin{equation} \label{radyator}
\sigma(s)=\sigma_{SM}(s)+\int^{1-4m_{\mu}^2/s}_0\ud xH(x)\left(\sigma_U[s(1-x)]+\sigma_{\rm int}[s(1-x)]\right)\,,
\end{equation}
where \cite{sigmaref}
\begin{equation}
H(x)=\beta x^{\beta-1}\delta^V+\delta^h
\end{equation}
with
\begin{eqnarray}
\beta&=&\frac{2\alpha}{\pi}(L-1)\,,\quad L=\log\frac{s}{m^2_e}\,,\quad
\delta^V=1+\frac{\alpha}{\pi}\left(\frac32L+\frac{\pi^2}{3}-2\right)+\ldots\,,\nonumber\\
\delta^h&=&\frac{\alpha}{\pi}(L-1)(x-2)+\ldots\,.
\end{eqnarray}
The LEP1-Aleph data are fitted with full QED corrections, since
the corrections to $\sigma_{SM}$ are not removed.

The corrected FBA for KEK-Venus and PETRA-MarkJ data is obtained in a similar manner \cite{fbaref}:
\begin{equation}
{\rm FBA}(s)=\frac{1}{\sigma(s)}\left[\sigma^{FB}_{SM}(s)+\int_{4m_{\mu}^2/s}^1\ud z\frac{4z}{(1+z)^2}\tilde{H}(z)
(\sigma^{FB}_U(zs)+\sigma^{FB}_{\rm int}(zs))\right]\,,
\end{equation}
where
\begin{eqnarray}
\sigma^{FB}&=&\int_{\theta>\pi/2}\ud\Omega\frac{\ud\sigma}{\ud\Omega}-\int_{\theta<\pi/2}\ud\Omega\frac{\ud\sigma}{\ud\Omega}\,,\\
\tilde{H}(z)&=&H(1-z)+\left(\frac{\alpha}{2\pi}\right)^2L^2\bigg[\frac{(1-z)^3}{2z}-(1+z)\log (z)+2(1-z)\nonumber\\
&&+\frac{(1-z)^2}{\sqrt{z}}\left(\arctan\frac{1}{\sqrt{z}}-\arctan\sqrt{z}\right)\bigg]\,.
\end{eqnarray}
Again, the LEP1-Aleph data are fitted with full QED corrections.

At the energy scale $M_Z$ with the scaling breaking parameter $\mu\gtrsim1$
GeV, unparticle  bremsstrahlung is not effective and thus not
included.

\clearpage
\providecommand{\href}[2]{#2}\begingroup\raggedright\endgroup

\end{document}